\documentclass[10pt,conference]{IEEEtran}
\usepackage{cite}
\usepackage{amsmath,amssymb,amsfonts}
\usepackage{graphicx}
\usepackage{textcomp}
\usepackage{xcolor}
\usepackage{xspace}
\usepackage{stfloats} 
\usepackage{color}
\usepackage{bm}
\usepackage{cite}
\usepackage{algpseudocode}
\usepackage{algorithm}
\usepackage[algo2e]{algorithm2e} 
\usepackage{setspace}
\usepackage{subfigure}
\usepackage[left=0.57in,right=1.59cm, top=.75in, bottom=1.01in]{geometry} 
\allowdisplaybreaks  

\long\def\comment#1{}

\newtheorem{defn}{Definition}

\def \snrtx{{\mathrm{SNR}_{\rm tx}}}

\let\bkslsh\backslash

\let\mbf\mathbf

\newcommand{\signf}{significant\xspace}

\newcommand{\env}{environment\xspace}

\newcommand{\sumimo}{SU-MIMO\xspace} 
\newcommand{\compatb}{compatible\xspace}

\newcommand{\matlab}{MATLAB\xspace}

\newcommand{\Unconsd}{Unconstrained\xspace}
\newcommand{\unconsd}{unconstrained\xspace}

\newcommand{\wifi}{WiFi\xspace}

\newcommand{\bsl}{baseline\xspace}

\newcommand{\freq}{frequency\xspace}

\newcommand{\corrtn}{correlation\xspace}

\newcommand{\tx}{transmitter\xspace}

\newcommand{\subcr}{subcarrier\xspace}
\newcommand{\subcrs}{subcarriers\xspace}
\newcommand{\ortho}{orthogonal\xspace}
\newcommand{\orthoty}{orthogonality\xspace}

\newcommand{\bfm}{beamforming\xspace}

\newcommand{\SemiOrtho}{Semi-Orthogonality\xspace}

\newcommand{\semiortho}{semi-orthogonality\xspace}

\newcommand{\zfbf}{ZFBF\xspace}

\newcommand{\ofdma}{OFDMA\xspace}
\newcommand{\mimo}{MIMO\xspace}
\newcommand{\mumimo}{MU-MIMO\xspace}
\newcommand{\pros}{ProxySelect\xspace}

\newcommand{\req}{requirement\xspace}

\newcommand{\coeffts}{coefficients\xspace}
\newcommand{\coefft}{coefficient\xspace}

\newcommand{\optn}{optimization\xspace}

\newcommand{\ie}{i.e.\xspace}

\newcommand{\Ip}{In particular\xspace}
\newcommand{\af}{as follows\xspace}
\newcommand{\resp}{respectively\xspace}

\newcommand{\etal}{\textit{et al.}\xspace}

\newcommand{\Fex}{For example\xspace}

\newcommand{\indeply}{independently\xspace}

\newcommand{\algo}{algorithm\xspace}

\newcommand{\indiv}{individual\xspace}

\newfont{\bbb}{msbm10 scaled 700}

\newfont{\bb}{msbm10 scaled 1100}

\newcommand{\hv}{{\mathbf{h}}}

\newcommand{\rv}{{\mathbf{r}}}

\newcommand{\wv}{{\mathbf{w}}}

\newcommand{\xv}{{\mathbf{x}}}

\newcommand{\ith}{{$i^{\rm th}$ }}

\newcommand{\kth}{{$k^{\rm th}$ }}
\newcommand{\nth}{{$n^{\rm th}$ }}

\newcommand{\lth}{{$\ell^{\rm th}$ }}

\newcommand{\Cc}{{\cal C}}

\newcommand{\Fc}{{\cal F}}
\newcommand{\Gc}{{\cal G}}

\newcommand{\Kc}{{\cal K}}

\newcommand{\Nc}{{\cal N}}

\newcommand{\Rc}{{\cal R}}

\newcommand{\Tc}{{\cal T}}

\newcommand{\diag}{{\hbox{diag}}}

\newcommand{\eqdef}{\stackrel{\Delta}{=}}

\newcommand{\be}{\begin{equation}}
\newcommand{\ee}{\end{equation}}
\newcommand{\bea}{\begin{eqnarray}}
\newcommand{\eea}{\end{eqnarray}}

\def\BibTeX{{\rm B\kern-.05em{\sc i\kern-.025em b}\kern-.08em
    T\kern-.1667em\lower.7ex\hbox{E}\kern-.125emX}}

\SetKwInput{KwData}{Input}
\SetKwInput{KwResult}{Output}
\RestyleAlgo{ruled}
\SetKwComment{Comment}{\%}{}

\begin{document}
\title{ProxySelect: Frequency Selectivity-Aware Scheduling for Joint OFDMA and MU-MIMO in 802.11ax WiFi}
\author{\IEEEauthorblockN{Xiang Zhang\IEEEauthorrefmark{1}, 
Michail Palaiologos\IEEEauthorrefmark{2},
Christian Bluemm\IEEEauthorrefmark{2}, 
and Giuseppe Caire\IEEEauthorrefmark{1}
}
\IEEEauthorblockA{
Department of Electrical Engineering and Computer Science, Technical University of Berlin\IEEEauthorrefmark{1}\\
Huawei Technologies Duesseldorf GmbH, Munich, Germany\IEEEauthorrefmark{2}\\
Email:~\IEEEauthorrefmark{1}\{xiang.zhang, caire\}@tu-berlin.de,~\IEEEauthorrefmark{2}\{michail.palaiologos1, christian.bluemm\}@huawei.com
}
}

\maketitle

\begin{abstract}
IEEE 802.11ax introduces orthogonal frequency division multiple access (OFDMA) to WiFi  to support concurrent transmissions to a larger number of users. As bandwidth continues to grow, WiFi channels exhibit increased frequency selectivity, which poses new challenges for MU-MIMO user selection: the optimal user set varies across frequency and is interleaved over subbands (called resource units, or RUs). This frequency selectivity, coupled with the complex subband allocation pattern, renders conventional narrowband user selection algorithms inefficient for 802.11ax. In this paper, we propose \emph{ProxySelect}, a scalable and frequency selectivity-aware user scheduling algorithm for
joint OFDMA and MU-MIMO usage in 802.11ax under zero-forcing beamforming (ZFBF). 
The scheduling task is formulated as an integer linear program (ILP) with binary variables indicating user (group)-RU associations,  and linear constraints ensuring standard compatibility. 
To reduce complexity, we introduce a novel proxy rate--a function of individual channel strengths and their correlations--that approximates the ZFBF rate without requiring cubic-complexity matrix inversion.
Additionally, we develop a sampling-based candidate group generation scheme that selects up to $T$ near-orthogonal user groups for each RU, thereby bounding the ILP size and ensuring scalability.
Simulations using realistic ray-tracing-based channel models show that ProxySelect achieves near-optimal rate performance with significantly lower complexity. 
\end{abstract}

\section{Introduction}
\label{sec:intro}
To meet the growing demand for higher data rates and improved efficiency in future WiFi networks, multiuser (MU) transmission has become a key component of the IEEE 802.11ax standard (also known as WiFi 6)~\cite{9442429}.
Specifically, with respect to downlink transmission, an access point (AP), equipped with multiple antennas, can leverage the spatial degrees-of-freedom to achieve  high user sum rates with \mumimo transmission. In addition, by employing \ofdma, users can be multiplexed in the frequency domain, thus leading to increased transmission efficiency. To achieve this, the available bandwidth is partitioned into subbands containing various numbers of consecutive \subcrs, called resource units (RUs).
To combine the benefits of both schemes, joint \ofdma and \mumimo transmission is introduced in 802.11ax, with the goal of optimally allocating RUs to user \emph{groups}. This problem is notoriously difficult to solve in practice though. Specifically, due to the combinatorial nature of the user group selection problem in \mumimo, the search space grows exponentially and becomes prohibitively large even with a moderate number of users~\cite{yoo2006optimality, dimic2005downlink}. Besides, user selection has to account for the frequency-selective nature of \wifi channels and, thus, needs to be optimized for each RU. As RUs with different sizes can be configured in the 802.11ax \ofdma frame, the optimal association between user groups and RUs depends on the RU configuration within the frame too, thus further complicating the problem~\cite{8422767, wang2018scheduling}.

As a result, only a few studies have considered the joint optimization of \ofdma and \mumimo for 802.11ax. For example,  Wang and Psounis~\cite{wang2018scheduling} proposed a divide-and-conquer algorithm, by leveraging the SIEVE user selection algorithm~\cite{shen2015sieve} for narrowband  channels. As this proposal is not fully compliant with 802.11ax specifications, they also proposed a suboptimal recursive algorithm and a sequential greedy \algo. In~\cite{lee2019using}, the design of a novel \ofdma uplink frame was leveraged towards optimizing the downlink \mumimo transmission. An exhaustive search over user groups of different sizes is required to identify the optimal user group, which is impractical due to its intractable complexity. 
A novel resource allocation scheme, based on deep learning, was proposed in~\cite{9448323}, although a fixed RU configuration of the \ofdma frame was used in the corresponding optimization, thus limiting the performance. 
Noh \etal~\cite{10534284} proposed a deep reinforcement learning-based approach for joint \ofdma and \mumimo  optimization. However, it is unclear how  the proposed approach can effectively navigate through the vast combinatorial search space while adhering to the complex user-RU association rules.
Overall, efficient user scheduling under joint \ofdma and \mumimo remains an open and challenging problem.

In this paper, we also focus on the joint OFDMA and MU-MIMO resource allocation problem in 802.11ax, with the aim of maximizing the downlink sum rate. 
We adopt zero-forcing beamforming (ZFBF) due to its effectiveness in eliminating multiuser interference~\cite{caire2003achievable,yoo2006optimality}. However, computing the beamforming vectors under ZFBF requires inverting the channel matrix, which can impose a substantial computational burden on the AP, especially in wideband systems exhibiting strong frequency selectivity. 
To address this challenge, we evaluate the degree of \emph{semi-orthogonality} among users within the RUs and derive a novel \emph{proxy rate} that approximates the true ZFBF rate without requiring matrix inversion. The proxy rate is computationally efficient and serves as the objective function in an integer linear program (ILP), where binary variables represent user or user-group assignments to RUs, and the constraints enforce compliance with the frequency and spatial domain requirements of the 802.11ax standard, \ie,  each user may be assigned to at most one RU, and MU-MIMO is permitted only on RUs that contain 106 \subcrs or more.
To further reduce complexity, we propose a sampling-based candidate group generation method that constructs up to $T$ distinct semi-orthogonal user groups for each RU. These groups are then fed into the ILP to obtain the optimal user scheduling. The parameter $T$ effectively controls the ILP’s complexity--specifically, the number of optimization variables and constraints--thereby ensuring scalability.
We evaluate the performance of ProxySelect using realistic, ray-tracing-based channel models in a large indoor conference room scenario. Experimental results demonstrate that ProxySelect achieves higher sum-rate performance than several existing baseline schemes, while also incurring significantly lower complexity. Throughout the paper, we denote $[N]\eqdef\{1,\cdots,N\}$.

\section{System Model and Problem Statement}
\label{sec: system model and problem description}
\begin{figure}[t]
    \centering
    \includegraphics[width=0.42\textwidth, height=0.16\textwidth]{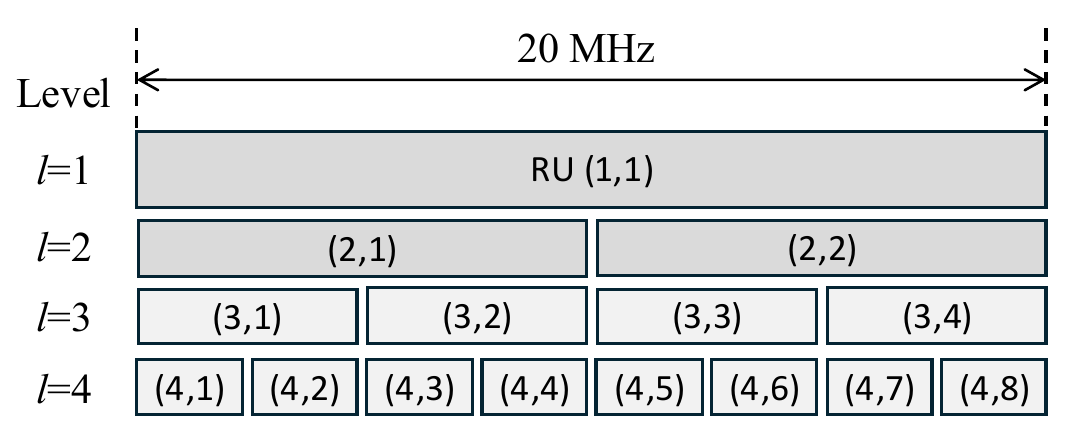}
    \vspace{-.4cm}
    \caption{OFDMA resource units for the 20 MHz channel. \mumimo is only allowed on RUs at levels 1 and 2 (shaded). }
    \label{fig: RU tree, 20 MHz}
    \vspace{-.5cm}
\end{figure}
Consider a WiFi access point (AP) with $N_T$  antennas, $K$ single-antenna users (STAs) and a total of $W\in  \{20,40,80,160\}$ MHz of bandwidth according to the 802.11ax standard~\cite{9442429}. 
We assume $K \geq N_T$ to fully exploit the multiplexing gain offered by \mumimo. 
Let $\hv_k[n]\in \mathbb{C}^{N_T}$ denote the \freq-domain channel vector of the \kth user on the \nth subcarrier/tone. 
Each subcarrier is $78.125$ kHz over which the channels are assumed to be flat. Channels may vary across subcarriers due to frequency-selective fading.
Also denote $\mbf{H}[n] \eqdef \left[\hv_1[n], \hv_2[n],\cdots, \hv_K[n]\right] \in \mathbb{C}^{N_T \times K} $ as the channel matrix  of all users on subcarrier $n$. For brevity  of notation,  the subcarrier index $n$ will be omitted whenever it causes no confusion. 
Let $s_k, \wv_k\in \mathbb{C}^{N_T}$ and $P_k$ denote \resp the data symbol, \bfm vector, and transmit power scaling factor for user $k$. The downlink transmitted signal is $\xv=\sum_{k=1}^K \sqrt{P_k} \wv_k s_k $ and user $k$ receives
\be
\label{eq: y_k formula}
y_k =\big( \sqrt{P_k}\hv_k^T \wv_k  \big )s_k+
\sum\nolimits_{i\ne k } \sqrt{P_i}\hv_k^T \wv_is_i+z_k
\ee 
where $z_k\sim \Cc\Nc(0,1)   $ is the complex AWGN noise. User $k$ detects the data symbol  $s_k$ by treating interference as noise, resulting in the rate
\be 
\label{eq: BF rate}
R =\sum\nolimits_{k=1}^K \log_2\left(  \frac{1+\sum_{i=1}^K P_i|\hv_k^T\wv_i|^2 } { 1+\sum_{ i=1, i\ne k   }P_i|\hv_k^T\wv_i|^2  }  \right).
\ee

\subsection{User Selection under ZFBF}
\label{subsec: ZFBF intro}
In ZFBF~\cite{caire2003achievable}, the \bfm vectors $\mathbf{W}\eqdef [\wv_1,\cdots, \wv_K] \in \mathbb{C}^{N_T\times K}   $  are chosen such that  the interference at unintended users is nullified, \ie, $\hv_k^T\wv_i=0,\forall i\ne k$. Suppose $\Kc \eqdef \{k_1, \cdots, k_{|\Kc|}\}$ is a subset of no more than $N_T$ users scheduled by the AP, and
 $\mbf{H}_{\Kc}\eqdef [\hv_{k_1}, \cdots, \hv_{k_{|\Kc|}}] \in \mathbb{C}^{N_T\times |\Kc|}  $ and $\mbf{W}_{\Kc}\eqdef [\wv_{k_1}, \cdots, \wv_{k_{|\Kc|}}] \in \mathbb{C}^{N_T\times |\Kc|}  $ denote \resp  the channels and \bfm vectors of these  users. A popular choice of $\mathbf{W}_{\Kc}$ is the pseudoinverse of $\mathbf{H}_{\Kc}$, \ie, 
\be
\label{eq: ZF beamformer}
\mathbf{W}_{\Kc} = \mbf{H}_{\Kc}^\dagger=  \mathbf{H}_{\Kc}^H\left( \mathbf{H}_{\Kc}\mathbf{H}_{\Kc}^H  \right)^{-1}.
\ee 
Then, (\ref{eq: BF rate}) becomes
\be
\label{eq: ZF rate}
R_{\textrm{ZFBF}}(\Kc) = \sum\nolimits_{k\in \Kc}\log_2\big(1+  \widetilde{ P}_k||\hv_k||^2 \big)
\ee 
where $\widetilde{P}_k\eqdef ||\wv_k||^2P_k$ denotes the per-stream transmit power.
As a common practice  in current WiFi products, we assume a constant per-stream power of $P$ across users, \ie, $\widetilde{P}_k=P,\forall k$. Because the noise has unit variance,  the  \tx-side signal-to-noise ratio (SNR) is equal to $\snrtx=P$.

\subsection{User Scheduling in 802.11ax}
\label{subsec: RU structure}
802.11ax introduced \ofdma to \wifi which  enables the joint use of \ofdma and \mumimo,  supporting more users in the same time. The entire bandwidth is partitioned into multiple resource units (RUs), each comprising a number of consecutive subcarriers (tones). Specifically, 
depending on the total bandwidth ($20/40/80/160$ MHz),  the RU partition forms a binary tree
with $L\in  \{4,5,6,7\} $ levels where the \lth  level consists of  $2^{\ell-1
}$ equally-sized RUs each containing a block of $26\times 2^{L-\ell}$ tones. There are $Q\eqdef 2^{L}-1$ different RUs in total. The smallest RUs--these at the bottom layer  of the RU tree--contains 26 tones each.
Denote RU $(\ell,i)$ as the \ith  RU, from left to right, in the \lth level. 
The RUs can also  be represented by a scalar index $q\in[Q] $ through the  bijection map $q$ $\Leftrightarrow$  $(\ell,i)$ where $\ell= \lceil \log_2(q+1)\rceil,i =q+1- 2^{\ell-1  } $.
The RU partition on the 20 MHz channel is shown in Fig.~\ref{fig: RU tree, 20 MHz} as an example.

Due to  OFDMA, different subsets of users can be scheduled on non-overlapping RUs free of interference. Therefore, user scheduling  in 802.11ax
consists of \emph{two  tasks}: \emph{i)  Partition the whole bandwidth into a number of RUs}. For example, a valid RU partition in Fig.~\ref{fig: RU tree, 20 MHz}  is $ \{(2,1), (3,3),  (4,7), (4,8)\}$.
\emph{ii) Assign a user or a user group to each RU}. These two tasks must be jointly performed, subject to  the following constraints: 1) {Each user can be assigned to at most one RU}; 2) {SU-MIMO mode: Allowed on all RUs. For RUs containing less than 106 tones, only \sumimo is allowed}; 3) {MU-MIMO mode: For RUs containing at least 106 tones, a group of at most $N_T$ users  can be scheduled}. 
For the 20 MHz channel, MU-MIMO mode is only allowed on RUs (1,1), (2,1) and (2,2). 

\begin{figure}[ht]
    \centering
    \includegraphics[width=0.28\textwidth, height=0.18\textwidth]{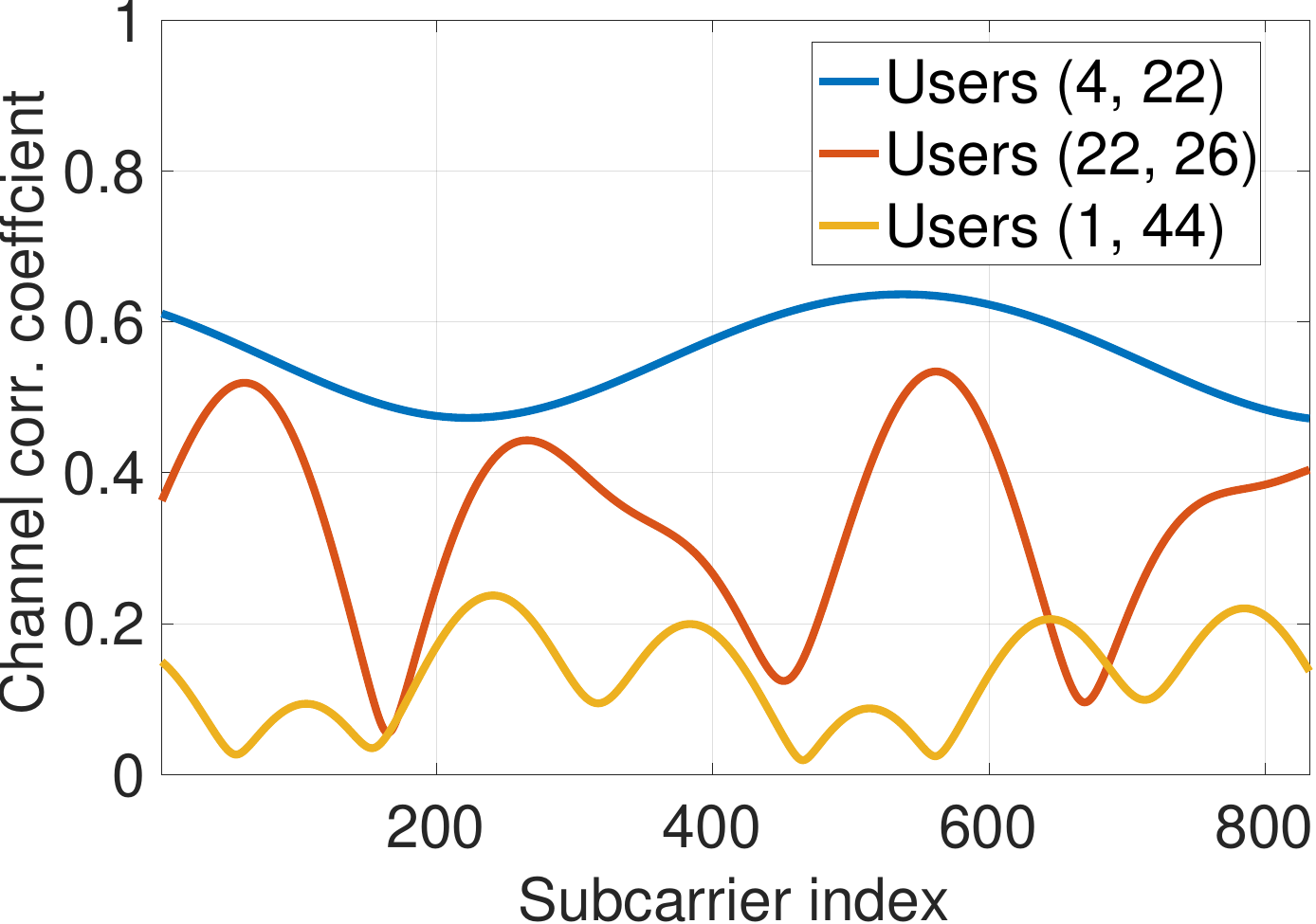}
    \label{fig:alpha vs freq}
     \vspace{-0.3cm}
    \caption{Channel correlation versus frequency.}
    \vspace{-0.3cm}
\end{figure}

User scheduling in 802.11ax is challenging primarily because of two factors. First, narrowband \mimo  user selection itself is hard due to the  exponentially large search space $\binom{K}{N_T}$, which becomes prohibitive when $K$ and/or $N_T$ is large~\cite{yoo2006optimality, dimic2005downlink}. In addition, evaluating the rate (\ref{eq: ZF rate})  under \zfbf  requires channel matrix inversion  with $O(N_T^3)$ cubic complexity per \subcr, which is cumbersome with large $N_T$ and many \subcrs.  
Second, as the available bandwidth becomes larger,  \wifi channels exhibit  \signf  \freq selectivity. For example, Fig.~\ref{fig:alpha vs freq} shows the pairwise channel correlation coefficient $\alpha$ of several pairs of users over the 80 MHz channel.
\vspace{-.2cm}
It can be seen that there is significant fluctuation in $\alpha$ as frequency changes. For example, users 22 and 26 are nearly-\ortho on some \subcrs but highly correlated on others. 
Since the \zfbf rate depends heavily on
$\alpha$, the optimal user group can vary across RUs.
Therefore, the \req that each user may be assigned to at most one RU couples user selection across different RUs, further complicating the problem.


\section{Proposed Approach}
\label{sec:proposed approach} 
This  section presents \pros, the proposed \freq selectivity-aware and scalable user scheduling algorithm for 802.11ax. In \pros, user scheduling is first formulated as an integer linear program (ILP) with binary \optn variables $x_q^\tau\in  \{0,1\}$ indicating whether a  user group $\tau\subset [K]$  is scheduled on RU $q$, and constraints enforcing the 802.11ax association rules. To reduce the number of variables, which corresponds to the candidate user groups to be considered in the ILP, a prescreening of the user groups is applied based on the principle of \semiortho developed by Yoo and Goldsmith~\cite{yoo2006optimality}. It  is known that the \zfbf rate $R_{\rm ZFBF}(\Kc)$ (See (\ref{eq: ZF rate}))  is maximized, without compromising the individual user's SNR, if the channels of users in $\Kc$ are mutually \ortho. Based on this observation, \cite{yoo2006optimality} proposed to select users with nearly \ortho channels (referred to as \semiortho). We also adopt this  principle. \Ip,  we first define \emph{$\alpha$-compatibility} to measure the extent of \semiortho among users within a group on a single subcarrier and on a RU which contains multiple subcarriers. Users with poor \orthoty will be prescreened and not considered in the ILP optimization. 
Based on $\alpha$-compatibility, we  derive a novel \emph{proxy rate} $R^{\rm proxy}(\Kc)$ to replace  the actual \zfbf rate
$R_{\rm ZFBF}(\Kc)$ used in the ILP objective, thereby
circumventing channel matrix inversion when evaluating the group rate. The proxy rate relies only on the \indiv channel strengths in $\Kc$, and an upper bound $\alpha\in [0,1]$ on their pairwise channel correlation, which can be computed efficiently. Empirical evidence shows that, with a reasonable choice of $\alpha$, the proposed proxy rate closely approximates the true \zfbf rate, incurring minor performance loss. The  impact of $\alpha$ will be discussed in Section~\ref{subsec: impact of alpha and T}.
Finally, to generate $\alpha$-compatible groups in a scalable manner, we adopt a sampling-based candidate group generation procedure. This procedure begins with a randomly selected user and incrementally adds new users who are compatible with all previously selected members, and stops if no remaining users are compatible. The sampling process is repeated $T$ times, resulting in up to $T$ distinct candidate groups for each RU. Consequently, the optimization variables can be defined for the ILP, solving which gives the optimal user scheduling.

To summarize, \pros comprises four essential components: an ILP formulation, a novel proxy rate, a user prescreening method based on channel semi-\orthoty, and a scalable sampling-based candidate group generation method. We describe each of these components in the following.

\begin{defn}[$\alpha$-Compatibility]
\label{defn: alpha compatibility}
For any two users $i,j$, their channel correlation coefficient on subcarrier $n$ is defined as 
\be
\label{eq: def alpha_ij[n]}
\alpha_{i,j}[n]\eqdef \frac{|\hv_i[n]^H \hv_j[n]   |   } {\|\hv_i[n]\|\|\hv_j[n]\| }\in [0,1].
\ee
A user group $ \Kc \subset [K] $ is said to be $\alpha$-compatible (on subcarrier $n$) if $  \alpha_{i,j}[n] \le \alpha,\forall i,j \in \Kc$.
\end{defn}

\begin{defn}[$\alpha$-Compatibility on RUs]
\label{defn: alpha compatibility on RU q} 
The average correlation between two users $i,j$ on RU $q\in [Q]$ is defined as
\be 
\label{eq: def alpha_ij avg on RU q}
\overline{\alpha}_{i,j} \eqdef ({1}/{|\Fc_q|})\sum\nolimits_{n\in \Fc_q} \alpha_{i,j}[n]
\ee 
where $\Fc_q$ denotes the  \subcrs contained in RU $q$.
A group $\Kc $ is said to be 
$\alpha$-compatible on RU $q$ if $\overline{\alpha}_{i,j} \le  \alpha,\forall  i,j\in \Kc$.
\end{defn}

\subsection{Derivation of the Proxy Rate}
\label{subsec: derivation of proxy rate}
One major limitation of the  existing  \mimo  user selection algorithms under ZFBF~\cite{dimic2005downlink, wang2008user, shen2015sieve,wang2018scheduling} is that they require channel matrix inversion (See (\ref{eq: ZF beamformer})) when evaluating the sum rate. This operation typically incurs a  complexity of $O(FN_T^3)$, where $F$ is the total number of \subcrs. Such complexity can become prohibitive when either $N_T$ or $F$ is large. 
To reduce complexity, we propose a lightweight proxy rate  that closely approximates the ZFBF rate (\ref{eq: ZF rate}) while
avoiding the costly pseudoinverse computation. Given an $\alpha$-compatible user group $\Kc$, the proxy rate depends only on the \indiv channel strengths $\{\|\hv_k\|^2\}_{k\in  \Kc}$, $\alpha$ and the size of $\Kc$. The derivation of the proxy rate is described \af.

Suppose $\Kc$ is $\alpha$-compatible where $|\Kc|\le N_T$.  For subcarrier $n$,
the normalized ZF \bfm  matrix is given by $\widetilde{\mathbf{W}}_{\Kc}[n] =
\mathbf{H}^{\dagger}_{\Kc}[n] \mathbf{D}_{\Kc}^{ -1/2 }[n]$
where  $\mbf{H}_{\Kc}^\dagger[n] =\mathbf{H}_{\Kc}^H[n]\left( \mathbf{H}_{\Kc}[n]\mathbf{H}_{\Kc}^H[n]  \right)^{-1}  $ denotes the pseudoinverse  of $\mbf{H}_{\Kc}[n]$, and
$\mbf{D}_{\Kc}[n] =\mathrm{diag}(( \mbf{H}_{\Kc}^H[n]\mbf{H}_{\Kc} [n])^{-1})$ is a $|\Kc|\times |\Kc|$ diagonal matrix. 
To simplify notation, we omit the \subcr  index $n$ in what follows.
It can be shown that $\widetilde{\mathbf{W}}_{\Kc} \eqdef [\widetilde{\wv}_k]_{k\in  \Kc}  $ has the property that $\widetilde{\wv}_k$ is the unitary vector corresponding to the \ortho projection of $\hv_k$ onto the \ortho complement of the span of the other channels $\{ \hv_i\}_{i\in \Kc \backslash \{k\}  }$.
Let $\mbf{H}_{\Kc \backslash k  }$ denote the matrix $\mbf{H}_{\Kc}$ after deleting  $\hv_k$. Then, the \ortho projector  for generating $\widetilde{\wv}_k$ is 
\be
\label{eq: ZF gen ortho projector}
\mbf{P}_{\Kc\backslash k}  = \mbf{I}_{N_T}
- \mbf{H}_{\Kc \backslash k }  \big( \mbf{H}_{\Kc \backslash k }^H  \mbf{H}_{\Kc \backslash k }  \big)^{-1}\mbf{H}_{\Kc \backslash k }^H 
\ee 
and 
$\widetilde{\wv}_k= \frac{\mbf{P}_{\Kc\backslash k} \hv_k  } {\|\mbf{P}_{\Kc\backslash k} \hv_k\|}, k \in \Kc$.
The useful signal coefficient for user $k$ (See (\ref{eq: y_k formula})) is therefore given by 
$\hv_k^H \widetilde{\wv}_k  =  \frac{\hv_k^H  \mbf{P}_{\Kc\backslash k}   \hv_k    } {\|\mbf{P}_{\Kc\backslash k}   \hv_k \|} =\|\mbf{P}_{\Kc\backslash k}   \hv_k \|$,  
where we used  the fact that \ortho projectors are Hermitian symmetric and idempotent, \ie, $\mbf{P}_{\Kc\backslash k}^H \mbf{P}_{\Kc\backslash k}=\mbf{P}_{\Kc\backslash k} $. As a result, the rate achieved by user $k$ on \subcr $n$  is equal to
\begin{align}
\label{eq:form 1, indiv rate in Kc}
  & \log_2\left(1+ \|\mbf{P}_{\Kc  \backslash k } \hv_k  \|^2 \snrtx  \right)=\notag\\
  & 
 \hspace{.8cm}\log_2\left( 1+ \hv_k^H \mbf{P}_{\Kc \bkslsh k}  \hv_k  \snrtx   \right).
\end{align}

We now bound the quantity $\hv_k^H \mbf{P}_{\Kc \bkslsh k} \hv_k$ in (\ref{eq:form 1, indiv rate in Kc}) in terms of the \indiv channel strengths $\{\|\hv_k\|^2\}_{k\in \Kc}$, the maximum correlation \coefft $\alpha$, and the size of the group $|\Kc|$. First, we can write the target quantity as
\begin{align}
\label{eq:form 1, useful signal component}
 \hv_k^{H} \mbf{P}_{\Kc\bkslsh k } \hv_k&= 
\hv_k^{H} 
\big( \mbf{I} - \mbf{H}_{\Kc\bkslsh k  } \big(\mbf{H}_{\Kc\bkslsh k}^{H} \mbf{H}_{\Kc\bkslsh k}\big)^{-1} \mbf{H}_{\Kc\bkslsh k}^{H} \big) \hv_k \notag \\
&= \|\hv_k\|^2 - \hv_k^{H} \mbf{H}_{\Kc\bkslsh k} \big(\mbf{H}_{\Kc\bkslsh k}^{H} \mbf{H}_{\Kc\bkslsh k} \big   )^{-1} 
 \mbf{H}_{\Kc\bkslsh k}^{H} \hv_k 
 \notag \\
&= \|\hv_k\|^2 - \rv_k^{H} \mbf{G}_{\Kc\bkslsh k}^{-1} \rv_k
\end{align}
where $ \rv_k \eqdef\mbf{H}_{\Kc\bkslsh k}^{H} \hv_k$ denotes the $(|\Kc|-1)\times 1$ vector containing the \corrtn \coeffts between $\hv_k$ and every other $\hv_{i}, i \in  \Kc\bkslsh \{k\}$. $ \mbf{G}_{\Kc\bkslsh k} \eqdef \mbf{H}_{\Kc\bkslsh k}^{H} \mbf{H}_{\Kc\bkslsh k}      $ denotes the $(|\Kc|-1)\times (|\Kc|-1)   $ Gram matrix containing the inner products  $\hv_i^H \hv_j$ for all $i,j\in \Kc\bkslsh \{k\}$. A simple form for $\mbf{G}_{\Kc \bkslsh k}$ is when all vectors have and identical real inner product $\alpha$ so that $\rv_k = (\alpha,\cdots, \alpha)^T$ and
\be 
\label{eq: G approximation}
\mbf{G}_{\Kc\bkslsh k}   = \mbf{D}_{\Kc\bkslsh k}  \left( (1 - \alpha) \mbf{I} + \alpha \mbf{1} \mbf{1}^{T} \right) \mbf{D}_{\Kc\bkslsh k} 
\ee 
where $\mbf{D}_{\Kc\bkslsh k} \eqdef  \diag(\{\|\hv_i\|^2\}_{i \in \Kc  \bkslsh \{k\} }  )   $  contains the magnitudes of the channels. Then the term
$\rv_k^H \mbf{G}_{ \Kc \setminus k  }^{-1}\rv_k $ in (\ref{eq:form 1, useful signal component}) is equal to 
$
\rv_k^H \mbf{G}_{ \Kc \setminus k  }^{-1}\rv_k 
=\alpha^2\|\hv_k\|^2 \bm {1}^T\left( (1-\alpha)\mbf{I} + \alpha \bm{1} \bm{1}^T \right )^{-1} \bm{1}
$. Using the Sherman-Morrison matrix inversion lemma~\cite{sherman1950adjustment}, with some tedious but
straightforward algebra, we obtain
$  
\rv_k^H \mbf{G}_{ \Kc \setminus k  }^{-1}\rv_k
= \alpha^2\|\hv_k\|^2 \frac{|\Kc|-1}{1+\alpha(|\Kc|-2)} 
$. Plugging this back into (\ref{eq:form 1, useful signal component}), we have
$  \hv_k^{H} \mbf{P}_{\Kc\bkslsh k } \hv_k = \|\hv_k\|^2\big(1- \frac{\alpha^2(|\Kc|-1)}{1+\alpha(|\Kc|-2)}\big)$. 
As a result, we obtain a proxy rate, as an approximation of the \zfbf rate for user group $\Kc$ on \subcr  $n$, given by 
\be 
\label{eq: proxy rate}
\sum\nolimits_{k\in \Kc }\log_2\left(1+\snrtx  \|\hv_k[n]\|^2\left(1- \frac{\alpha^2( |\Kc|-1)}{1+\alpha(|\Kc|-2)}\right)\right).\notag
\ee 

In the proposed proxy rate, we get rid of matrix inversion and the group rate is expressed simply in terms of \indiv channel strengths and  $\alpha$, which can be easily computed. It is worth mentioning that  when either $\alpha=0$ or $|\Kc|=1,2$,  the  proxy rate is \emph{exactly} the same as the true \zfbf rate.  Empirical evidence suggests that, with a reasonably small $\alpha \le 0.3$, the proxy rate can closely mimic the true \zfbf rate. This means that using the proxy rate as the ILP objective does not incur notable performance loss  if  $\alpha$  is  relatively small.
However, small $\alpha$ might be too restrictive that we cannot select enough number of \compatb groups to feed to the ILP, leading to suboptimal solutions.  Therefore, the choice  of $\alpha$ can be optimized experimentally
depending on the specific propagation \env, which will be discussed in Section~\ref{subsec: impact of alpha and T}.

\subsection{Integer Linear Programming Formulation}
\label{subsec: integer LP formulation}
Given a suitable average channel correlation score $ \alpha$, the set of compatible user groups  can be  determined for each RU. \Ip, let  $ \Tc_q \subseteq 2^{[K]} \backslash \{\emptyset\}  $ denote the set of all $\alpha$-compatible groups on RU $q$ where $T_q \eqdef  |\Tc_q|,q \in[Q]$. Note that $T_q\ge K$ because a single user is always compatible on any RU.
For each $q$ and $\tau \in \Tc_q$, a binary  variable  $x^\tau_q\in \{0,1\}$ is defined to indicate if user group $\tau$ is assigned to RU $q$. 
Denote 
\begin{align}
& R_k^{\rm proxy}(\tau,q) \eqdef\notag  \\
& \quad \sum\nolimits_{n\in \Fc_q} \log_2 \left(1+ \snrtx \|\hv_k[n]\|^2 \left( 1- \frac{\alpha^2(|\tau|-1)}{1+ \alpha (|\tau|-2)} \right) \right)\notag
\end{align}
as the proxy rate for user $k \in \tau  $  when group $\tau $ is scheduled on RU $q \in [Q]$.
The sum rate maximization problem is then formulated as 
\begin{subequations}
\label{eq: OPT original, mumimo ofdma}
\begin{align}
 & \max_{
\{x^\tau_q\}_{q\in[Q], \tau \in \Tc_q}
 }  
\sum_{q\in[Q]}\sum_{\tau \in \Tc_q}
  x^\tau_q \sum_{k\in  \tau} R_k^{\rm proxy}(\tau,q)\label{eq: obj, original}
\\
& \qquad\; \mathrm{s.t.} \quad \; 
 \sum\nolimits_{q\in[Q], \tau \in \Tc_q^k  } x_q^\tau  \le 1, \; \forall k\in  [K] \label{eq: c1}  \\
 &\qquad \qquad\quad  \sum\nolimits_{\tau \in \Tc_q} x^\tau_q \le y_q, \;  \forall  q\in[Q]  \label{eq: c2}  \\
 &\qquad \qquad \quad \;\,  y_q +y_{q'}\le 1,\; \forall (q,q')\in \Cc_{\rm RU} \label{eq: c3}
\end{align}
\end{subequations}
where $\Tc_q^k \eqdef\{ \tau \in \Tc_q: k\in \tau  \} $ in (\ref{eq: c1})  denotes the  candidate groups in $\Tc_q$ that includes user $k$; $ \Cc_{\rm RU}$ denotes the (undirected) RU conflict graph in which  an edge $(q,q')$ exists if RU $q$ is either an ancestor or a descendant of RU $q'$, i.e., $\Fc_q \cap \Fc_{q'} \ne \emptyset$. \Fex, in Fig.~\ref{fig: RU tree, 20 MHz}, RU $(1,1)$ is in conflict with all other RUs since it occupies the entire bandwidth.
The constraints (\ref{eq: c1})-(\ref{eq: c3}) ensure  the obtained user scheduling is compliant with the 802.11ax standard. \Ip, (\ref{eq: c1}) ensures that any user is assigned to at most one RU. (\ref{eq: c2}) ensures that each RU is assigned at most one user group. Moreover, (\ref{eq: c3}) ensures that the scheduled RUs are not in conflict with each other, \ie, among the  set of RUs that share a nonempty set of \subcrs, at most one of them can be assigned with some user group. In addition, the \req that only \sumimo mode is allowed on  small RUs  can  be easily incorporated by setting  $\Tc_q =\{\{1\} ,  \cdots, \{K\} 
 \}  $. 
(\ref{eq: OPT original, mumimo ofdma}) is a binary ILP, which can be solved efficiently using commercial solvers such as Gurobi~\cite{anand2017comparative}.

The proposed ILP formulation allows for arbitrary selection of candidate user groups, making it highly flexible. Here we applied a prescreening of groups based on the $ \alpha$-compatibility criterion and $|\Tc_q|\le T$ candidate groups are considered for each RU, resulting in at most $QT$ variables for the ILP. Since $Q$ is a fixed number depending on the bandwidth, the choice of $T$ reflects a trade-off between performance and complexity.


\subsection{Sampling-based Candidate Group Generation}
\label{subsec: random sampling based candidate group selection} 
Before solving the ILP (\ref{eq: OPT original, mumimo ofdma}), we need to determine the set  of candidate groups $\Tc_1, \cdots, \Tc_Q$, which define the \optn variables $\{x_1^{\tau}\}_{\tau \in  \Tc_1}, \cdots, \{x_Q^{\tau}\}_{\tau \in  \Tc_Q}       $. 
For relatively large $K$, it is nearly impossible to enumerate all subsets of users and check if each of them  is $\alpha$-\compatb. 
To generate these candidate groups $\Tc_1, \cdots, \Tc_Q$ in a scalable manner, we use a sampling-based procedure that can produce a number of at most $T$  compatible groups. 
\Ip, the procedure starts with randomly selecting one user. It then looks at the remaining users and
select one that is compatible with the selected user. To diversify the sampling process (\ie, to prevent the same group from being sampled repeatedly), we apply a random permutation to the order in which the remaining users are inspected. New users are added until none of the remaining users are compatible, or the number of selected users reaches $N_T$. We repeat the above sampling process $T$ times for each RU so that at most $ T$  distinct candidate groups are generated for RU $q$. Note that $|\Tc_q|\ge K$ because a single user is always  compatible on any RU. 
For small RUs where only \sumimo is allowed, we simply set $\Tc_q=\{\{1\},\cdots, \{K\} \},\forall q \ge 2^{L-2}$.

\subsection{Complexity Analysis}
\label{subsec: complexity analysis}
The candidate group generation process has a complexity of $O(KTQN_T^3)$, which scales linearly with $K$.
The resulting ILP contains at most $QT=(2^L-1)T$  variables and $K+Q+(L-2)2^L+2 $ constraints--both linear in $K$.
Although the number of variables and constraints grows exponentially with $L$, this does not pose a significant computational burden in practice, as $L \in \{4,5,6,7\}$, yielding no more than $127T$ variables and $K+833$ constraints.
Experiments in Section~\ref{subsec: execution time analysis} show that with up to $T=1000$, the ILP  can be efficiently solved in sub-second time using the Gurobi solver~\cite{anand2017comparative}.

\section{Experiments }
\label{sec:numerical evaluation}
\begin{figure}[ht]
    \centering
    \includegraphics[width=0.3\textwidth, height=0.18\textwidth]{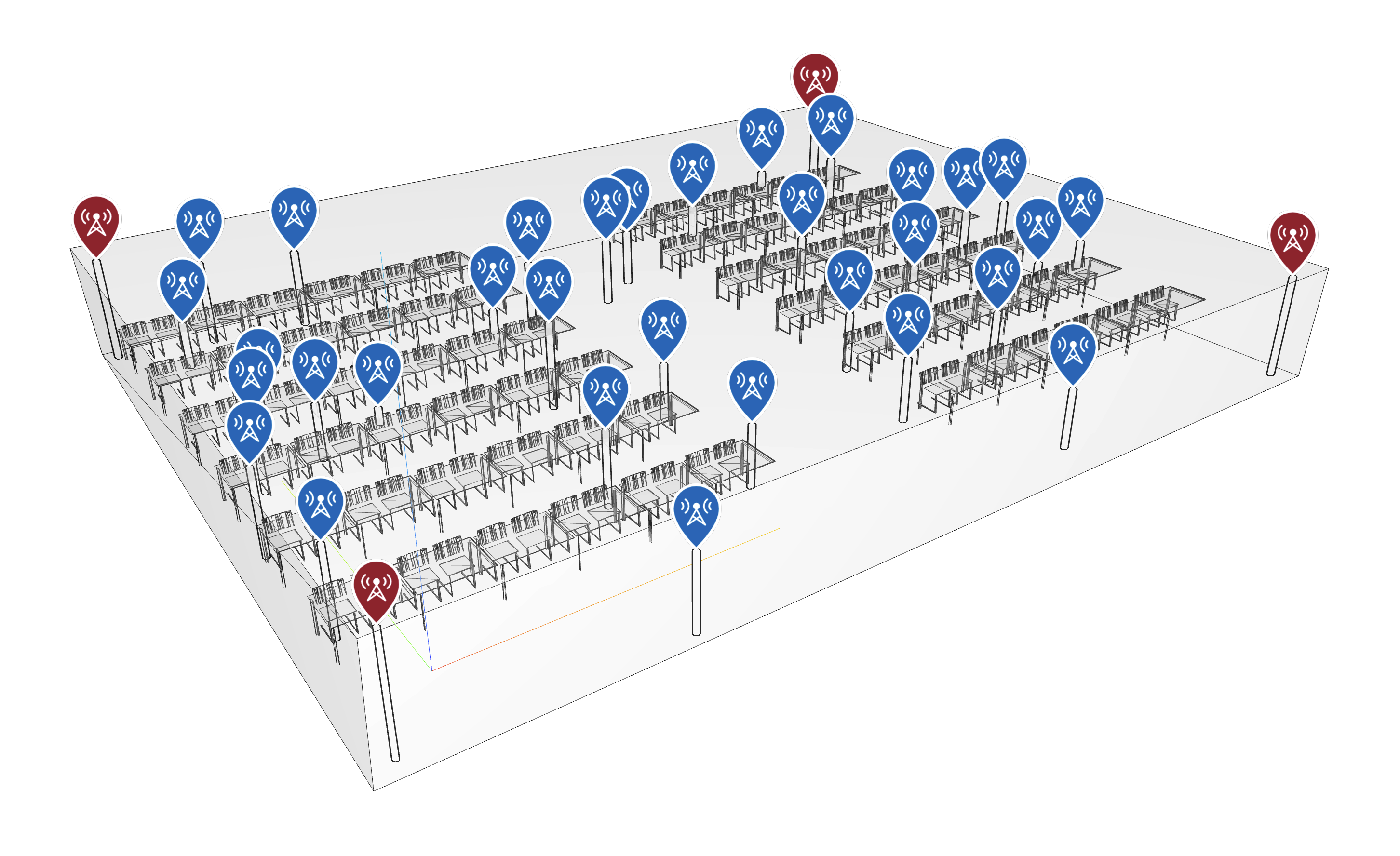}
    \vspace{-.5cm}
    \caption{Conference room with dimensions of $ 23\times 18 \times 2.8$ meters. 4 AP antenna heads (red) are placed at the corners at a height of $2.5$ meters. }
    \label{fig:conference room}
    \vspace{-.2cm}
\end{figure}

The proposed user scheduling algorithm is evaluated using realistic channels in a large conference room scenario as shown in Fig.~\ref{fig:conference room}. The channels are generated through ray-tracing using the $\mathrm{comm.RayTracingChannel}$ module of \matlab.    
The BSS consists of $K$ single-antenna users, and 4 AP antenna heads each having $4$ uniform linear array (ULA) antennas, placed at the corners of the room.
\vspace{-.4cm}
Thus, there are $N_T=16$ antennas in total. This setup is motivated by the fiber-to-the-room (FTTR) concept~\cite{Zhang2023} for future WiFi systems with distributed antenna heads and centralized processing. For each simulation scenario, we fix the location of the AP and generate 50 topologies by randomly distributing the users. The average sum rate over these topologies is then reported. Moreover,
for each topology, \pros is averaged over 10 random realizations of the sampling process.

\subsection{Baseline Schemes}
\label{subsec:baseline schemes}
We compare \pros with the following baselines:

\emph{1) Pure \ofdma}: At most one user can be scheduled on each RU. \mumimo is not used. The optimal assignment can be obtained by solving the proposed ILP with candidate groups $\Tc_q=\{\{1\},\cdots,\{K\} \},\forall q$.

\emph{2) Sequential greedy selection from~\cite{wang2018scheduling}}: First choose a level  $ 
\ell^* =\min\{L-2, \lfloor \log_2(K/N_T) \rfloor +1  \} 
$ 
to operate. Using the SIEVE algorithm~\cite{shen2015sieve} to select user groups for each RU in level $\ell^*$, from left to right, and exclude the previously chosen users when selecting for a new RU.

\emph{3) Greedy selection over whole bandwidth}: SIEVE selection~\cite{shen2015sieve} is used to select a user group across the entire bandwidth. 
\Ip, the algorithm begins by selecting the single user who, if scheduled on every \subcr, yields the highest sum rate. It then evaluates all possible pairs formed by combining this user with each remaining candidate, selecting the pair that maximizes the sum rate. This greedy process continues by incrementally adding users--each time choosing the one that, when added to the current group, provides the largest rate improvement--until the group reaches $\min\{N_T, K\}$ users. Finally, the algorithm reviews the sequence of user additions and selects the user group along this trajectory that achieves the highest overall rate.
This scheme has a complexity of $O(FKN_T^4)$, which can be slow with a larger number of \subcrs $F$ and/or antennas $N_T$.

\emph{4) \Unconsd greedy selection}:
SIEVE selection~\cite{shen2015sieve}  is performed \indeply on each \subcr, which may result in overlapping user groups that are not standard-compliant. Nevertheless, this \bsl serves as a performance upper bound.

\subsection{Impact of \SemiOrtho Threshold $\alpha$ and $T$}
\label{subsec: impact of alpha and T}
\begin{figure}[!ht]
    \centering
    \subfigure[\small Sum rate vs. $\alpha$]{     \includegraphics[width=0.223\textwidth, height=0.18\textwidth]{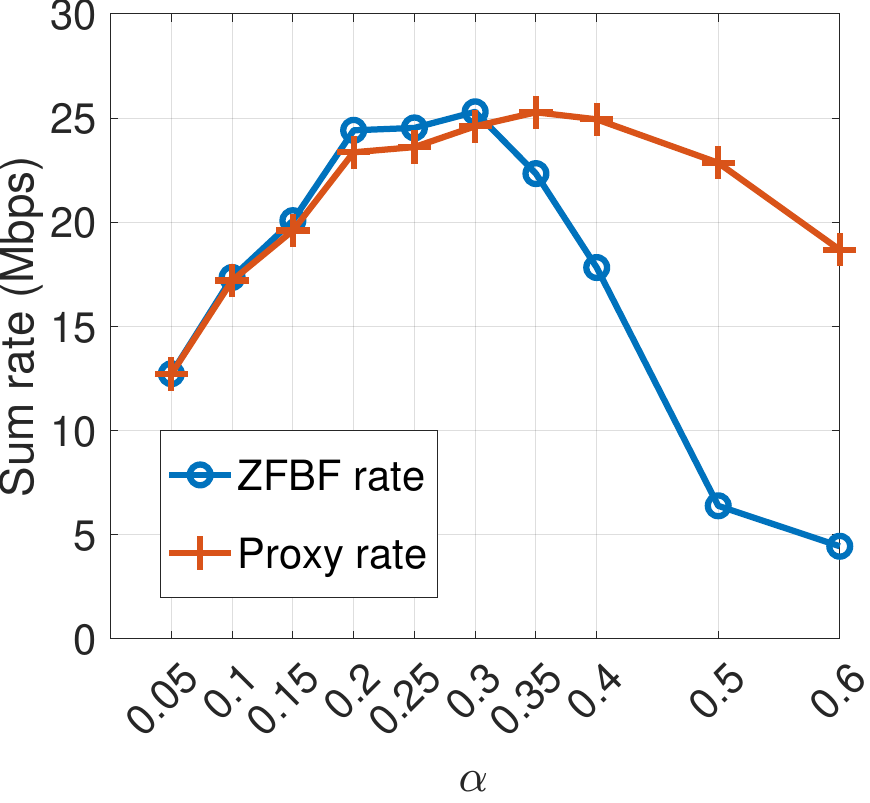}
    }
    \subfigure[\small Sum rate vs. $T$]{  \includegraphics[width=0.23\textwidth,height=0.18\textwidth]{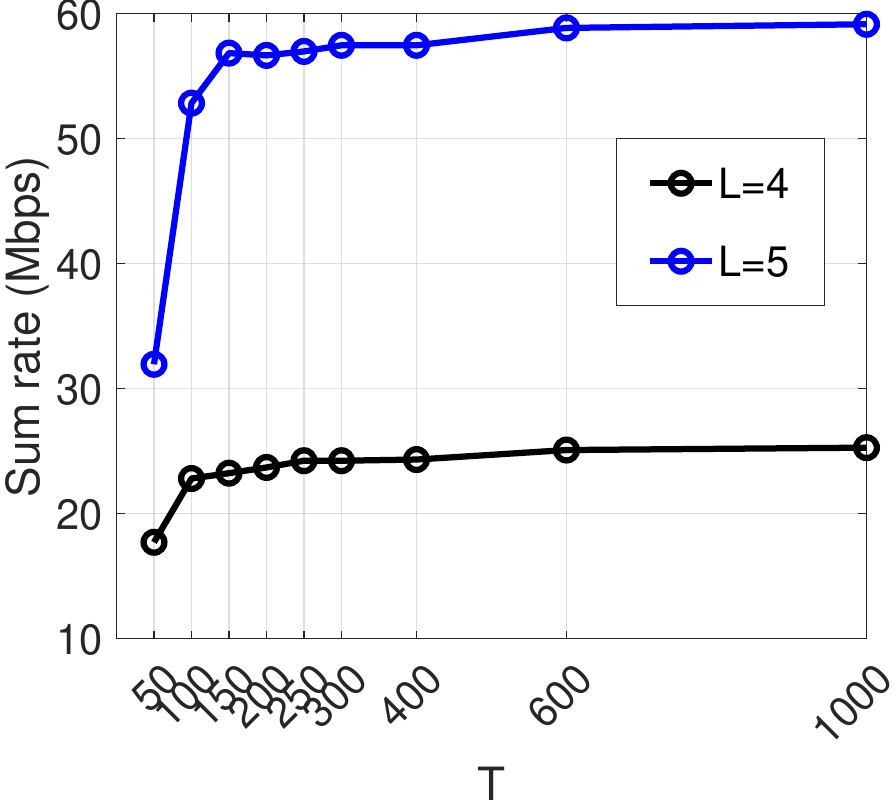}
    }
    \vspace{-.4cm}
    \caption{\small Impact of $\alpha$ and $T$.}
    \label{fig:impact of alpha and T}
    \vspace{-.3cm}
\end{figure}

As discussed earlier, the performance of \pros depends on the \semiortho threshold $\alpha$. To determine the best $\alpha$, we compare the proxy rate and the actual ZFBF rate produced by \pros, and choose  an $\alpha$ value that maximizes the \emph{actual} ZFBF rate.
Fig.~\ref{fig:impact of alpha and T} (a) depicts  the average proxy and ZFBF rates corresponding to various $\alpha$ values on the 20 MHz ($L=4$) channel with $K=48$ users.
\vspace{-.3cm}
It  can be seen that the proxy rate approximates the ZFBF rate quite well up until $\alpha=0.3$. When $\alpha>0.3$, the actual ZFBF rate begins to decrease although the proxy rate keeps increasing. This is because a large $\alpha$ increases the likelihood that user groups with poor \orthoty are selected during sampling, thereby degrading the actual ZFBF performance. For other bandwidth values, a similar turning point at $0.3$ can be observed. Therefore, a suitable  threshold for the considered propagation \env is $\alpha=0.3$.
In addition, Fig.~\ref{fig:impact of alpha and T} (b) shows the impact of the sampling times $T$ on the actual ZFBF sum rate. For both the 20 MHz ($L=4$) and 40 MHz ($L=5$) channels, the sum rate increases with $T$ when $T \leq 150$. As $T$ further increases, the growth in rate gradually slows and eventually saturates when $T \geq 600$. This is because when $T$ significantly exceeds the actual number of $\alpha$-\compatb user groups, most of the \compatb groups are already captured by the sampling process, and further increasing $T$ does not introduce additional groups that yield notable rate improvements. We choose $T=1000$ in our simulations.

\subsection{Comparison with Baselines}
\label{subsec: comparison with baselines}
\begin{figure}[!ht]
    \centering
    \subfigure[\small Average sum rate]{     \includegraphics[width=0.22\textwidth, height=0.18\textwidth]{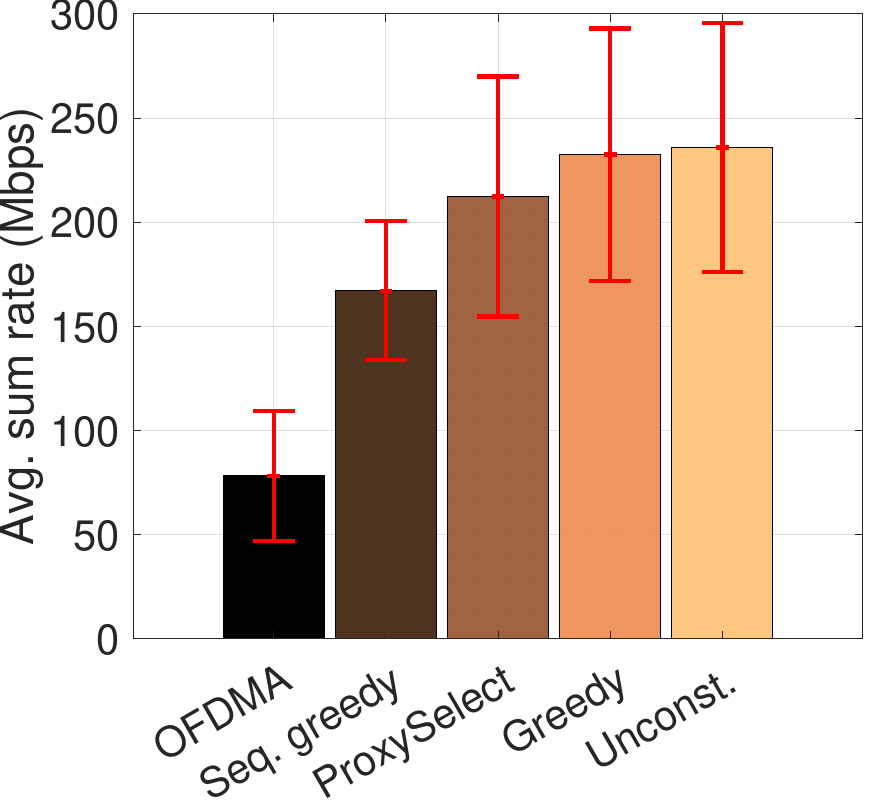}
    }
    \subfigure[\small Sum rate distribution]{  \includegraphics[width=0.23\textwidth, height=0.18\textwidth]{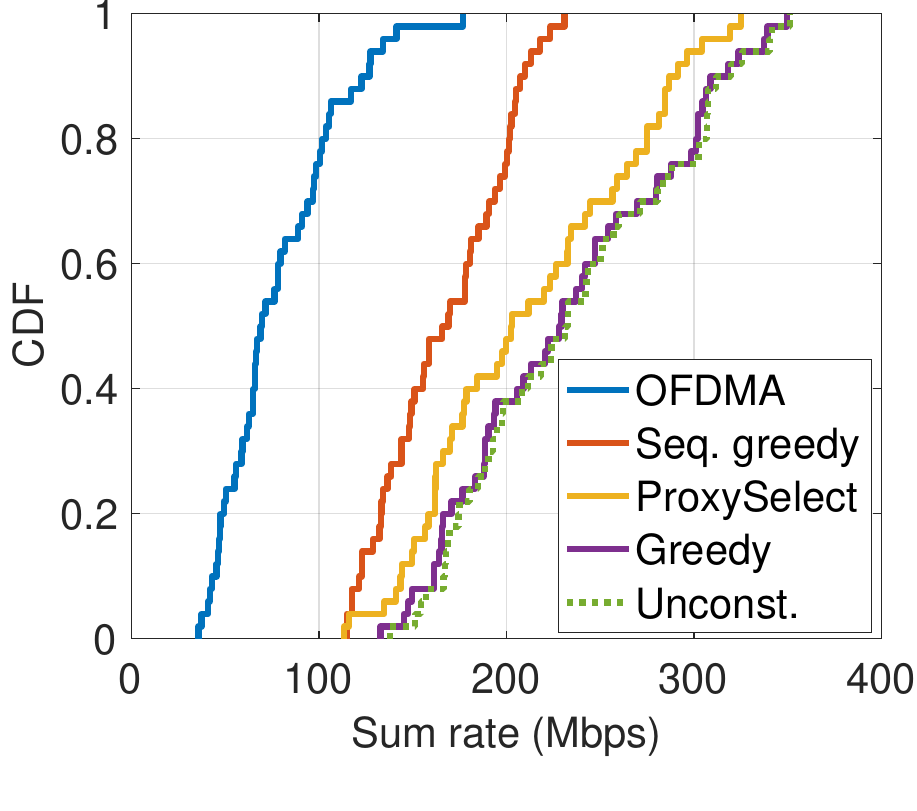}
    }
    \vspace{-.5cm}
    \caption{\small Comparison of sum rates.}
    \label{fig:comparison with baselines}
    \vspace{-.5cm}
\end{figure}

Fig.~\ref{fig:comparison with baselines} (a) shows the average sum rate of \pros and the baseline schemes over the 160 MHz bandwidth with $48$ users. Several observations emerge: First,
\pros delivers a $2.7\times$ rate improvement over pure \ofdma, highlighting the substantial benefit of combining \ofdma with \mumimo.
\pros also outperforms the sequential greedy algorithm by a significant margin, achieving a $27\% $  higher sum rate. Compared to the whole-bandwidth greedy selection, \pros achieves a slightly lower sum rate (within $9\%$). However, this marginal performance gain comes at a significant computational cost--as shown in the next section, \pros attains over $91\%$ of the rate while reducing execution time by a factor of four. In addition, \pros achieves $90\%$ of the rate of the unconstrained greedy selection, demonstrating its near-optimality, given that the \unconsd selection serves as a performance upper bound. The cumulative density function (CDF) of the achieved rates is shown in Fig.~\ref{fig:comparison with baselines} (b).
\vspace{-.3cm}

\subsection{Execution Time Analysis}
\label{subsec: execution time analysis}
\begin{figure}[t]
    \centering
    \subfigure[\small Exec. time vs. $L$]{     \includegraphics[width=0.22\textwidth, height=0.17\textwidth]{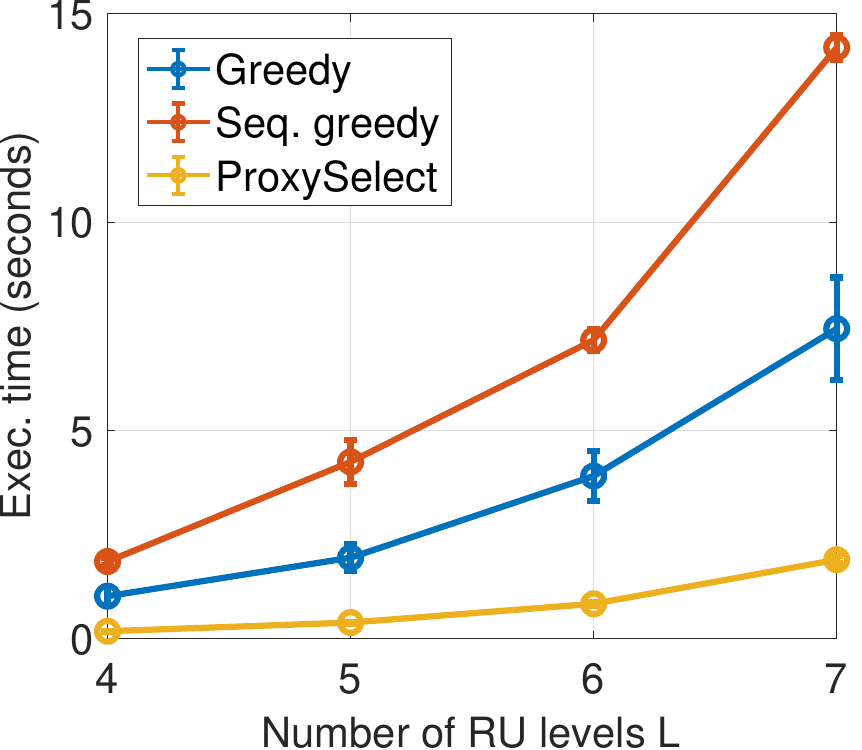}}
    \subfigure[\small Exec. time vs. $K$]{  \includegraphics[width=0.22\textwidth, height=0.17\textwidth]{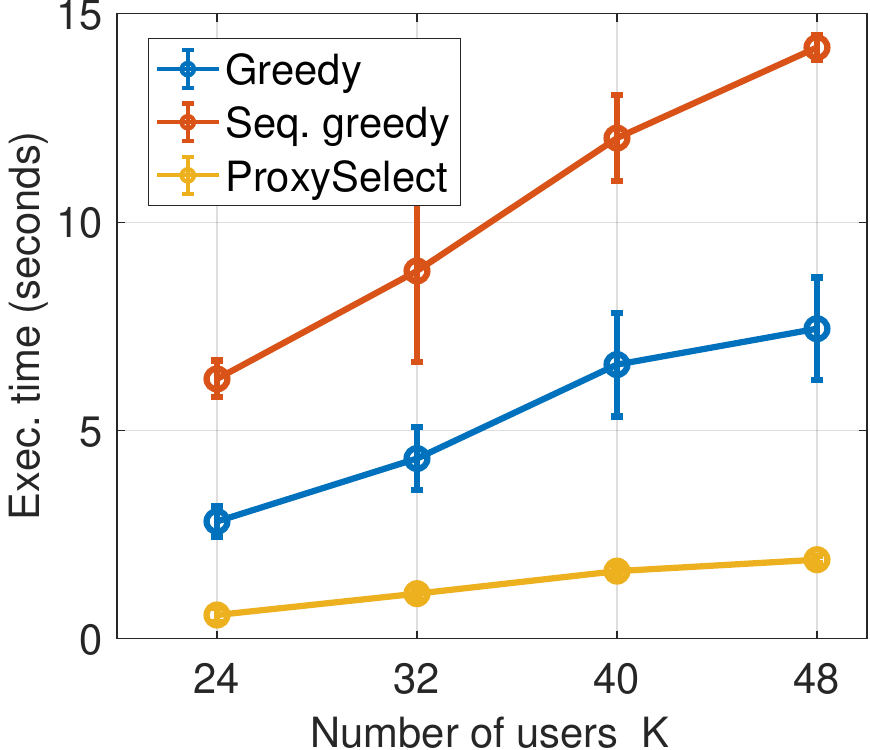}}
    \vspace{-.2cm}
    \caption{\small Comparison of execution time.}
    \label{fig: exe time comparison}
    \vspace{-.5cm}
\end{figure}

Fig.~\ref{fig: exe time comparison} shows the average algorithm execution time on a MacBook with Apple M3 chip and 16 GB RAM. As the bandwidth (equivalently, the number of RU levels $L$) and the number of users increase, the execution time also increases. However, \pros runs significantly faster than the two other baselines. For example, with 160 MHz $(L=7)$ bandwidth and $48$ users, \pros runs in 1.9 seconds, which is $4\times$ and $7\times$ times faster than the two baselines.  
The reduction in execution time is mainly attributed to both the use of the proxy rate, which avoids matrix inversion for rate evaluation, and the sampling-based candidate group generation procedure that controls the size of the ILP.
\vspace{-.2cm}

\section{Conclusion}
In this paper, we presented ProxySelect, a novel user scheduling algorithm designed for joint OFDMA and MU-MIMO optimization in 802.11ax WiFi. ProxySelect introduces a  proxy rate that closely approximates the true ZFBF rate while circumventing the computational overhead of matrix inversion. By leveraging $\alpha$-compatibility to assess user orthogonality and adopting a scalable, sampling-based candidate group generation method, ProxySelect formulates the scheduling problem as an integer linear program with manageable and tunable complexity.
Extensive simulations using realistic ray-tracing-based channel models demonstrate that ProxySelect can achieve near-optimal sum-rate performance with  reduced computational cost.  
These results highlight the effectiveness of combining channel orthogonality-based user prescreening with structured optimization, offering a practical and scalable solution for user  scheduling in frequency-selective WiFi environments.

\section{Acknowledgment}
The work of X. Zhang and G. Caire was supported by the Gottfried Wilhelm Leibniz-Preis 2021 of the German Science Foundation (DFG).

\bibliographystyle{IEEEtran}
\bibliography{references_wifi.bib}
\end{document}